# Blockchain platform for COVID-19 vaccine supply management

**Claudia Daniela Antal (Pop), Tudor Cioara, Marcel Antal, Ionut Anghel**

Computer Science Department, Technical University of Cluj-Napoca, Memorandumului 28, 400114 Cluj-Napoca, Romania; claudia.pop@cs.utcluj.ro; tudor.cioara@cs.utcluj.ro; marcel.antal@cs.utcluj.ro; ionut.anghel@cs.utcluj.ro

**Abstract:** In the context of the COVID-19 pandemic, the rapid roll out of a vaccine and the implementation of a worldwide immunization campaign is critical, but its success will depend on the availability of an operational and transparent distribution chain which can be audited by all relevant stakeholders. In this paper, we discuss how blockchain technology can be used for assuring the transparent tracing of COVID-19 vaccine registration, storage and delivery, and side effects self-reporting. We present such system implementation in which blockchain technology is used for assuring data integrity and immutability in case of beneficiary registration for vaccination, eliminating identity thefts and impersonations. Smart contracts are defined to monitor and track the proper vaccine distribution conditions against the safe handling rules defined by vaccine producers enabling the awareness of all network peers. For vaccine administration, a transparent and tamper-proof side effects self-reporting solution is provided considering person identification and administrated vaccine association. A prototype was implemented using the Ethereum test network, Ropsten, considering the COVID-19 vaccine distribution tracking conditions. The results obtained for each on-chain operation can be checked and validated on the Etherscan, demonstrating various aspects of the proposed system such as immunization actors and safe rules registration, vaccine tracking, and administration. In terms of throughput and scalability, the proposed blockchain system shows promising results.

## 1. Introduction

COVID-19 virus part of the coronavirus ribonucleic acid virus family [1] has generated a worldwide pandemic being very easy to spread and pushing a lot of pressure on the healthcare system and on levels of the society. Since its identification in Wuhan, China in December 2019, it has spread rapidly through community transmission generating up to December 2020 to around 65 million confirmed cases and more that 1.5 million deaths [2], [3]. Even if significant efforts have been made for fighting the pandemic, the spreading rate of the virus was only slowed. In many countries restriction measures still are still in place to avoid suffocating the hospitals and treatment centres [4].

In this context, the rapid roll out of a vaccine and the implementation of a worldwide immunization campaign is critical for the control of the pandemic. Since the beginning of the pandemic pharmaceutical companies have concentrated their efforts for developing a vaccine in record time to achieve COVID-19 containment [5], [6]. While some of COVID-19 vaccines are in the final test phases, preparing and planning for mass immunization becomes extremely important. Nevertheless, there are several aspects that are likely to affect the success of COVID-19 immunization program if they are not properly addressed.

The first aspect is the availability of an operational and transparent end-to-end supply chain and logistics systems [7], [8]. Its role is on one hand to assure the vaccine storage and stock management and on the other hand the rigorous temperature control in the cold chain [5]. Table 1 below show the storage and distribution requirements of the most likely vaccine candidates as they appear now.

**Table 1.** COVID-19 candidate vaccines temperature control conditions [10], [11].

|  | Freezer temperature | Refrigerator temperature | Max storage days |
|---|---|---|---|
| **Pfizer** | - 70 degrees Celsius | N/A | 30 days after opening the freezer |
| **Modena** | - 20 degrees Celsius | 2-8 degrees Celsius | 30 days in the refrigerator |
| **Oxford-AstraZeneca** | N/A | 2-8 degrees Celsius | 6 months in the refrigerator |

Blockchain can increase the efficiency and transparency of COVID-19 vaccine distribution assuring the traceability and the rigorous audit of the storage and delivery conditions. In our opinion, blockchain-based solutions may provide a fully automated implementation of data accountability and provenance tracking in vaccine distribution, which will enable the integration of different information silos as well owned and managed by different types of stakeholders on the entire distribution chain. Self-enforcing smart contracts may assure the traceability of the COVID-19 vaccine supply chain especially the cold part of the chain in which the vaccine needs to be kept at extremely low temperatures to remain viable. Moreover, a breach in assuring the delivery conditions will be registered on the chain in a tamper-proof manner and all the peers of the network will be made aware due to the distributed ledger block distribution and replication features. Furthermore, the blockchain can act as proof of the delivery chain, making it impossible to counterfeit the vaccine, since at any point the medical units and the vaccine beneficiaries would be able to trace it back up to the companies that have registered the vaccine lots in circulation.

The second aspect is the transparency and correctness in the registration and management of the waiting list of people for immunization. The data on this list is not only sensitive but at the same time, it requires correctness, avoidance of impersonation, privacy, and immutability. These properties can be achieved by using blockchain technology. Blockchain can change the way in which the waiting list is managed allowing parties mutually unknown to transact and trace digital assets securely without a central trusted intermediary. Such a decentralized system will remove the necessity of having third parties' entities that centralize and manage the waiting list the immutability of transactions and the authorization achieved by using smart contracts which allows all peers to restrict access to their private information. The entire sequence of actions taken in a smart contract may be propagated across the network and/or recorded on the blockchain, and therefore are publicly visible. Transactional privacy, as well as the privacy of personal data, can be assured using novel solutions such as the incorporation of zero-knowledge proofs which are cryptographic techniques that can assure privacy for verifying private data without revealing it in its clear form [12] [13].

Finally, the third aspect is building trust in vaccine by implementing a transparent and public reporting systems of potential side effects including the automatic tracing back up vaccine lot level. Concerns have raised that different drug makers do not correctly and completely report the side effects to relevant authorities [14] [15]. Thus, a transparent, real time and reliable system regarding the reporting of the side effects once a drug/vaccine is released is crucial. In this sense a blockchain platform would bring advantages with respect to the existing state of the art solutions. Any beneficiary that has received a vaccine, may report any problems/symptoms encountered after the administration using blockchain, all other peers will be made aware and moreover the report could be potentially validated using the peers' consensus in relation to vaccine lot. Furthermore, being stored in an immutable log, all the reported side effects are protected against tampering.

This paper introduces a blockchain-based system for transparent tracing of COVID-19 vaccine registration, storage and delivery and side effects self-reporting. It brings the following contributions:
- A blockchain based solution for data immutability, transparency and correctness of beneficiary registration for vaccination, eliminating identity thefts and impersonations;
- A decentralized smart contracts-based monitoring solution for assuring proper vaccine transportation conditions in a cold chain and real time awareness of all peers in relation to the fulfillment of COVID-19 vaccine delivery and storage conditions;
- A blockchain solution for vaccine administration and transparent and tamper proof self-reporting of side effects, person identification and vaccine association.

The rest of the paper is organized as follows: Section 2 describes the relevant related work in the area of Information and Communication Technology (ICT) and blockchain solutions for managing immunization campaigns; Section 3 presents the proposed blockchain system for safe vaccine distribution and Section 4 presents a test case for COVID-19 vaccine distribution tracking and system scalability results. Finally, Section 5 presents conclusions and future work.

**2. Related work**

ICT solutions for supporting immunization campaigns are proposed in the state of the art mostly for optimal distribution planning of vaccines [16], [17], [18]. In [19] a drive-through vaccination simulation tool is proposed for planning and feasibility assessment of such facilities based on event processing and agent-based modeling to minimize waiting times, staff required, immunization intervals, etc. Vaccine distribution for heterogeneous population has been approached by using mathematical modelling using equity constraint to maintain fairness and to optimize the number of vaccine doses in case of an influenza outbreak [17], [18]. Various heuristics and custom optimization algorithms are proposed for optimizing the distribution network design [20], [21], [22].

Recent advancements of contemporary technologies such as Internet of Things (IoT), machine learning and blockchain pave the way for building more smart and innovative systems that can be adapted to different domains as it is the case of the healthcare domain [23]. The authors of [24] propose the use of IoT devices to monitor the location of the carrier, temperature and humidity with the goal of optimizing and increasing vaccine coverage in the remote regions and ensuring transparency in the overall process. Blockchain based decentralized systems for addressing healthcare sector problems such as privacy and confidentiality of data are presented in [25], [26]. Recent studies have pointed the possibilities of using blockchain in combating the COVID-19 pandemic most of them addressing the decentralized tracking of contracts and symptoms or for assuring security and immutability [27]. Relevant use cases for blockchain technology in managing COVID-19 pandemic contact tracing, patient data sharing, supply chain management are overviewed in [28]. Other studies have shown that blockchain can be used to develop trustful predictive systems that can help containing the pandemic risks on national territory [29] or to securely track the movements of residents in quarantine scenarios using IoT infrastructures [30]. Incentive based approaches have been proposed to battle against the COVID-19 pandemic that use blockchain to prevent information tampering and incentives for rewarding patients to remain in quarantine [31].

Blockchain has been proposed as a solution for organization and management of industry supply chains [32]. For pharmaceutical supply chain where temperature monitoring or counterfeit drug prevention are of utmost importance, IoT and blockchain frameworks may offer a viable solution [33]. In [34] a blockchain drug supply chain management is combined with a machine learning recommendation system. The supply chain management system is deployed using Hyperledger fabrics to continuously monitor and track the drug delivery process while N-gram, LightGBM models are used to recommend the best medicines to the customers. In [35] Gcoin blockchain is proposed for the data flow of drugs to create transparent drug transaction data where every unit that is involved in the drug supply chain can participate in the same time to prevent counterfeit of drugs.

Few approaches in the literature are addressing the development of blockchain based systems for distribution of vaccines. Authors of [36] propose a blockchain management system for the supervision of vaccine supply chains through smart contracts also dealing with vaccine expiration and fraud recording. Machine learning models are used for recommendations to choose better immunization methods and vaccines. A blockchain model for the COVID-19 vaccine distribution chain is proposed in [37]. The approach allows monitoring each phase from development to application, considering the emerging and commercial chains while leveraging on blockchain for authenticating each process and registering changes. Similarly, in [38] efficient supply chain management through smart containers with IoT sensors is proposed and used to administer shipment, payments, legitimize receiver, etc. VeChain [39] is developing a blockchain-based platform for vaccine production and tracing in China using IoT devices to capture vaccine production data and store it an enterprise blockchain to ensure immutability. Finally, blockchain can help to track the vaccines and make sure they haven't been compromised or even to keep track of patients' vaccine records and provide proof of vaccination especially because COVID-19 will require two vaccine doses [40], [41].

## 3. Blockchain System for Vaccine Distribution

The proposed blockchain system for transparent COVID-19 vaccine tracking, distribution monitoring and administration is presented in Figure 1. It uses the distributed ledger for storing vaccine related data, while assuring data immutability to guarantee that the vaccines are transported safely to the beneficiaries and the administration is done correctly to the real recipient without abuses.

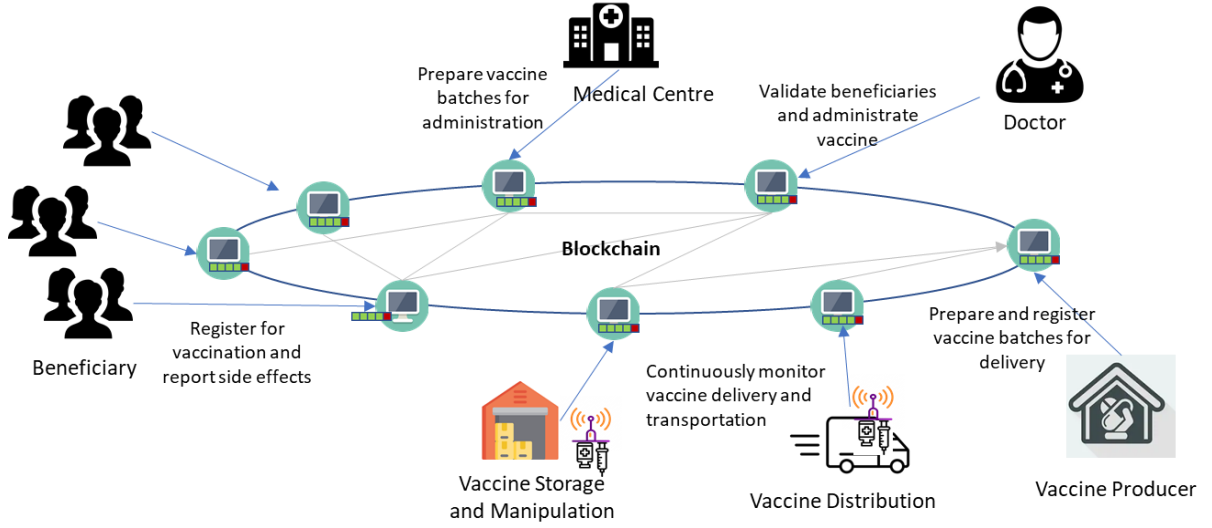

**Figure 1.** Blockchain and immunization program management: vaccine registration, tracking, monitoring, administration and self-reporting.

The main actors of the proposed blockchain based system that act as peer network nodes are: i) the beneficiaries that register for vaccination, ii) the company that prepares and registers the vaccine batches/lots for transportation, iii) the IoT sensor devices that continuously monitor the vaccine delivery, storage and handling; iv) the medical centers that will receive the vaccine and prepare it for administration and v) the doctor who validates the beneficiary, delivery and storage conditions and administers the vaccine. All the actions are registered into the distributed ledger as immutable transactions which are stored in blocks that are replicated to all the peer actors in the chain. This will provide a high transparency of the vaccine handling operations enabling the tracking and registration of the COVID-19 vaccine as digital asset.

The main features of the system as well as their implementation using self-enforcing smart contracts are detailed in the next sub-sections.

### 3.1. Immutable registration of vaccine beneficiaries

The objective of beneficiary registration for vaccination via the proposed blockchain based system is to ensure the privacy preserving of identity as well as to avoid the impersonations. As shown in Figure 2, before the actual registration, the beneficiary generates a secret key (SK) that will be stored off chain and it will be later used to prove his/her identity. To maintain personal data privacy and anonymity on chain, while at the same time enabling beneficiary to prove their identity without reveling it, a Merkle Proof [42] is used. The root node stores the hash of the secret key generated by the beneficiary and the hash of its Personal Identification Number (PI).

$$P\_HASH = Hash(Hash(PI), Hash(SK)) \qquad (1)$$

The root of the Merkle Proof ($P\_HASH$) becomes the payload of a blockchain transaction signed by the beneficiary and aimed for the Vaccine Registry ($V\_REG$) contract deployed on chain, marking the intent to receive the vaccine. Once mined, the transaction hash and the contract address are sent back to the beneficiary, who generates a QR Code that will be later shown as identification to the doctor containing the transaction hash, the contract address, the PI and the hash of the SK.

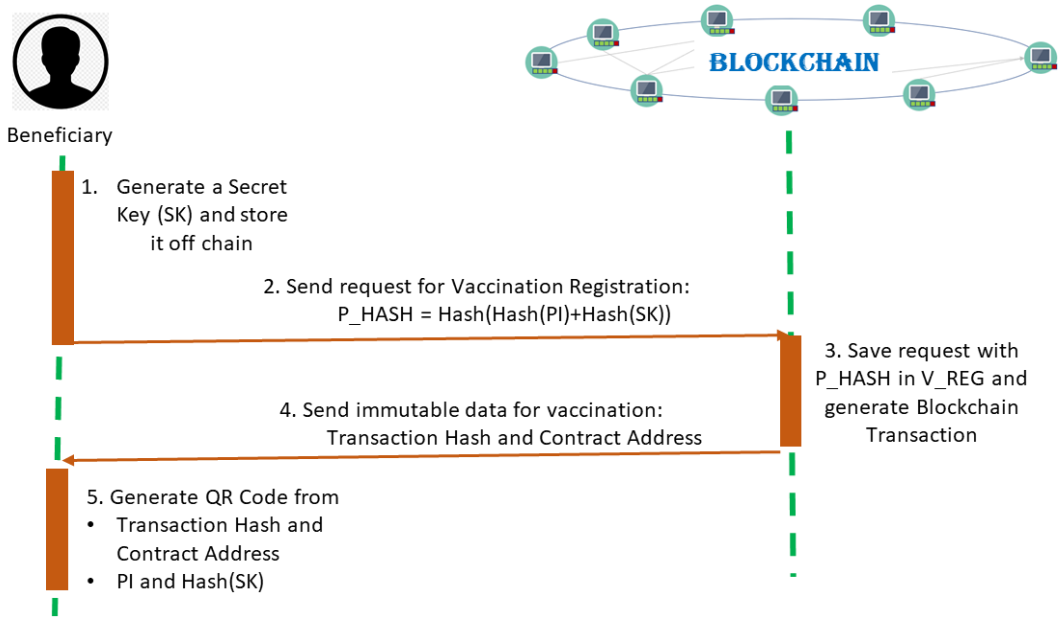

**Figure 2.** Immutable registration of vaccine beneficiary with the blockchain system.

All relevant actors' registration actions are management using smart contract functions (see Figure 3). We have used a map structure for keeping track of the registered actors, since it features a small time overhead for accessing the resources upon verification and validation. This leads to a smaller execution costs of the blockchain transactions mined. The beneficiary can register the request and intent for receiving the vaccine, and the address signing the registration transaction will be stored. As input payload, the beneficiary must provide the *beneficiaryHash* (line 10) represented by the Merkle Root. During the verification step conducted by the doctor, the beneficiary will need to reveal the raw data to prove that he/she is the actual person who made the registration request using a pseudo-anonymized blockchain address.

| **Smart Contract:** Actors Registration |
|---|
| 1: **State:** |
| 2:     **address** _vaccineIssuer |
| 3:     MAP (**address** doctor, **bool** inserted) _doctors |
| 4:     MAP (**address** admin, **bool** inserted) _medicalUnitAdmins |
| 5:     MAP (**address** beneficiary, **bool** inserted) _beneficiary |
| 6:     MAP (**bytes** beneficiaryHash, **address** beneficiary) _registeredRequests |
| 7: **Function Modifiers:** - onlyIssuer; – onlyMedicalManagers; - onlyFreezer |
| 8:     … |
| 9: **Function** *RegisterBeneficiary* |
| 10:     **Input:** msg.sender, beneficiaryHash |
| 11:     **Output:** - |
| 12:     **Modifiers:** - |
| 13:     **Begin:** |
| 14:         **_beneficiary** [msg.sender] ← **true** |
| 15:         **_registeredRequests** [beneficiaryHash] ← msg.sender |
| 16:     **End** |

**Figure 3.** Smart contract template for registering actors' actions on the chain.

Upon smart contract deployment, the vaccine issuer address is stored as being the address that signed the deployment transaction (line 2). The functions for actions registration associated with other actors such as Medical Centers, vaccine producers or IoT devices are surrounded by modifiers ensuring that their transactions are signed with corresponding verified addresses (line 7).

## 3.2. Vaccine distribution chain monitoring

The goal in this case is to make transparent the degree in which the defined conditions for vaccine storage and manipulation are met during the entire distribution chain. This is achieved using smart contracts that evaluate continuously the data received from sensors deployed on storage units or attached to the transportation freezers against the defined conditions rules (see Figure 4).

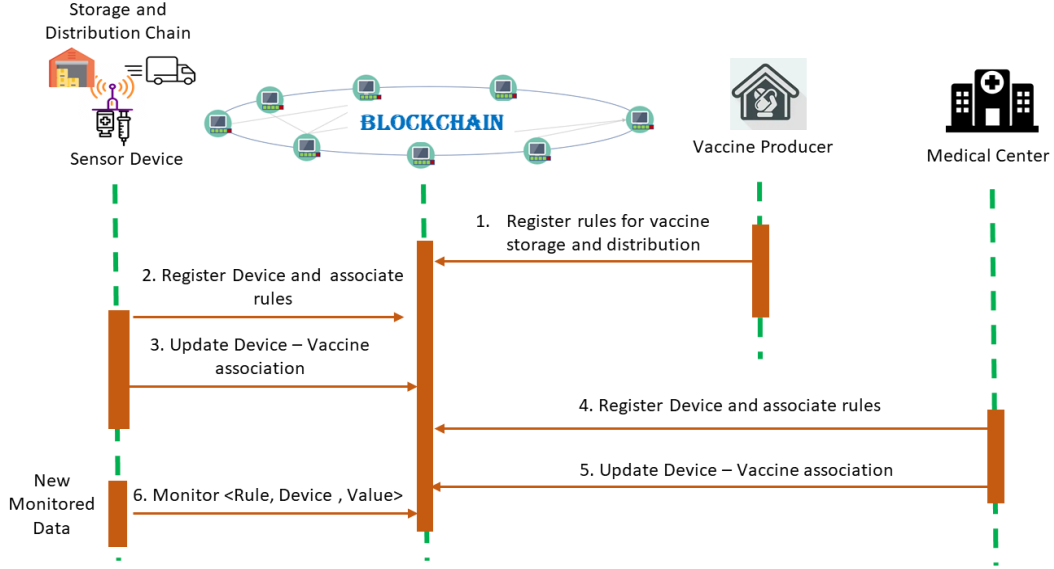

**Figure 4.** Monitoring the vaccine distribution chain.

First the vaccine producer company must register a set of R rules for safe distribution and storage of the vaccine batches (see Figure 4). The rules are encoded in smart contracts associated to specific IoT devices as rules that must be checked each time a new data is provided:

$$Rule_k: Min_{Value} < Monitored_{Value}(t) < Max_{Value}, \forall k \in \{1..R\} \text{ and } t < TIME_{LIMIT} \quad (2)$$

The companies in the distribution chain or the medical centers can register a set of freezing devices used for vaccine manipulation and storage, since the vaccine lots can be transported or stored in different type of devices depending on the destination distance and travel time (Figure 4). At the same time, they may update the freezing devices to vaccine correlations, by mapping the vaccine batch ready to be transported to a freezer ID that in his turn has associated a set of rules. All the association and rules are being stored in the blockchain distributed ledger making them impossible to be tampered.

The IoT devices are responsible to sign blockchain transactions containing the monitored data. It must contain the following payload: $< VL_{ID}, Rule_k, Monitored_{Value}(t) >$, specifying the monitored vaccine batch identifier $VL_{ID}$, the monitored value and the rule $Rule_k$. Once the transaction is mined, the contract will verify the identity of the device signing the transaction and the monitored value against the rule limits defined for safe vaccine handling. As the value reaches the blockchain, it triggers the computation of the smart contract rules defined validating or invalidating the transportation conditions. Furthermore, based on the time provided by the chain, the time limit imposed by the vaccine issuer regarding the transportation/storage may be validated:

By storing the monitored values and the rules on blockchain, the immutability and integrity of the data is assured. The monitored values cannot be changed and the decision of annotating these values as corresponding/breaking the issuer-imposed rules are subject to consensus and mined in chain in a tamper-proof manner. Any actor may check the logs registered on chain and trust that the results provided have not been subject to any malicious tampering.

Figure 5 presents the smart contract used to track the vaccine distribution against the defined handling and storage rules. The freezer devices and vaccine lots are registered on the chain. The vaccine lots are assigned to freezing devices, this association being updated during the distribution chain enacting its decentralized tracing (see lines 9-18). Each time a monitored value associated with a freezing device is provided it is checked against the imposed rules (line 26-29). The data regarding the monitored vaccine lot and all the information regarding the validity of the rule and the time of the registration are stored in the blockchain in a tamper-proof log managed by the contract (line 30-32).

---

**Smart Contract:** Vaccine Distribution Monitoring

```
1:   State:
2:       MAP (bytes vaccine_LID, int samples) _vaccineLots
3:       MAP (string rule, ImposedRules) _rules
4:       MAP (address freezer, MAP (bytes vaccine_LID, bool inserted) _freezers
5:       MAP (address freezer, MAP (bytes vaccine_LID, long time) _freezerRegistrationTime
6:       MAP (address freezer, MAP (string rule, bool inserted)) _freezerRules
7:       MAP (bytes vaccine_LID, MonitoredRule[]) _monitoredVaccines
8:       ….
9:   Function UpdateVaccineFreezer
10:      Input: msg.sender, block.timestamp, vaccineLotId, oldFreezer, newFreezer
11:      Output: -
12:      Modifiers: onlyMedicalManagers
13:      Begin:
14:          Requires _vaccineLots [vaccineLotId] to exist
15:          _freezers[oldFreezer] [vaccineLotId] ← false
16:          _freezers[newFreezer] [vaccineLotId] ← true
17:          _freezerRegistrationTime [newFreezer] [vaccineLotId] ← block.timestamp
18:      End
19:  Function Monitor
20:      Input: msg.sender, block.timestamp, vaccineLotId, rule, monitoredValue
21:      Output: -
22:      Modifiers: onlyFreezer
23:      Begin:
24:          Requires _vaccineLots [vaccineLotId] to exist
25:          Requires _freezers[msg.sender] [vaccineLotId] to exist
26:          Requires _rules [rule] to exist
27:          valid = _rules [rule].maxValue < monitoredValue &&
28:              _rules [rule].minValue > monitoredValue &&
29:              _rules [rule].timeDelta < (block.timestamp - _freezerRegistrationTime [msg.sender] [vaccineLotId])
30:          _monitoredVaccines[vaccineLotId].add(MonitoredValue(msg.sender, rule,
31:                                               monitoredValue, block.timestamp, valid)
32:      End
```

**Figure 5.** Smart contract for monitoring the vaccine distribution chain.

*3.3. Vaccine administration and side efects reporting*

The most complex operation of the pipeline is the actual vaccine administration. This step must check the following conditions for the blockchain system operation: the identity of the beneficiary to be vaccinated, the conditions of vaccine delivery and handling according to the rules defined by producer and the association of beneficiary with the vaccine to be administrated enabling the further reporting of potential side effects.

The first step of the process is the validation of beneficiary identity prior to the vaccination (see Figure 6). It is performed by the doctor using the beneficiary registration QR code containing the blockchain transaction hash, the smart contract address, the hash of personal identification number and the hash of the secret key. Using the hashes extracted from the QR code, the doctor performs an on-chain identity verification for registration acknowledgment. Using the Merkle Proof, the hash of the two values is compared against the root stored in the blockchain during the beneficiary registration step.

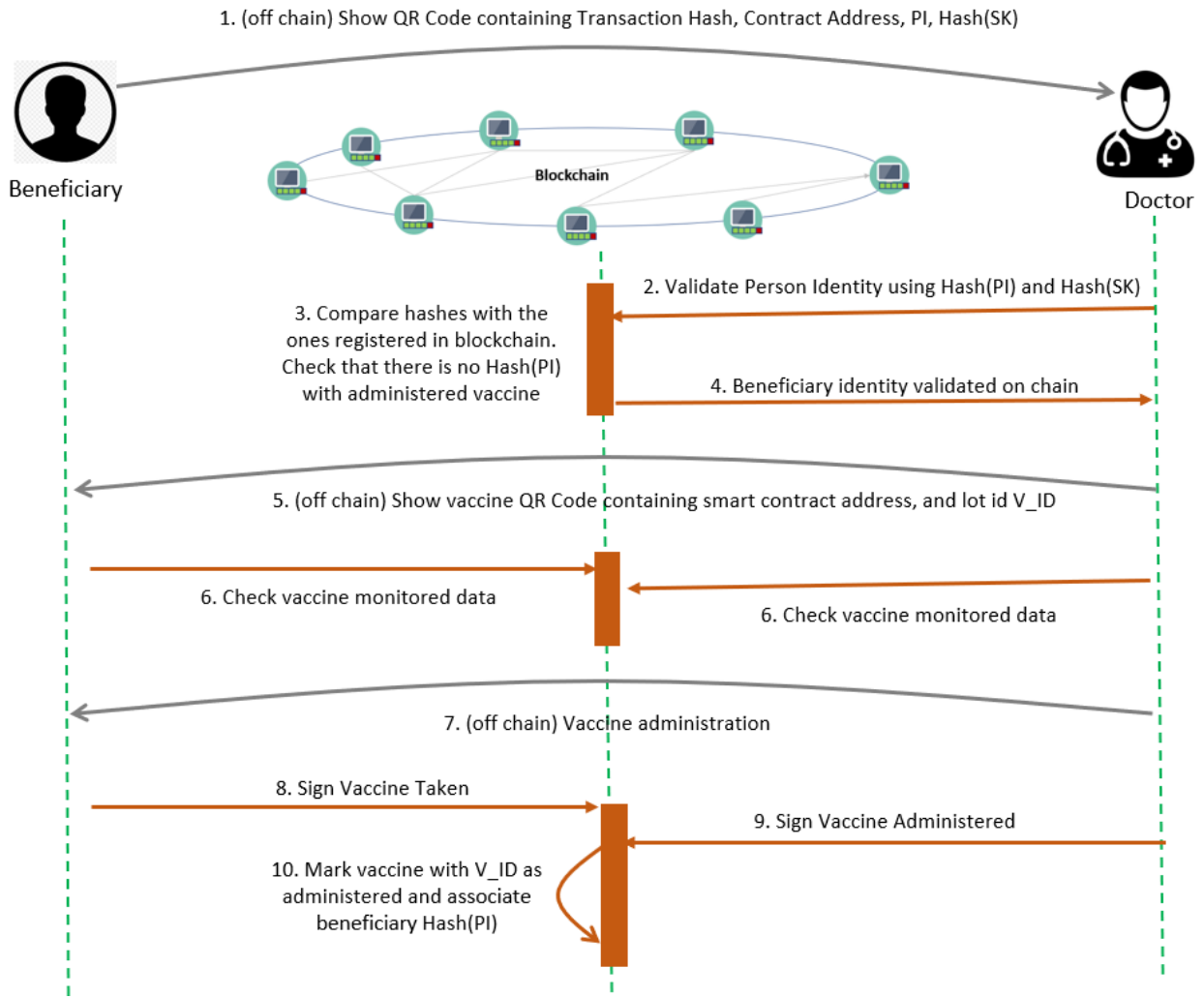

**Figure 6.** Vaccine Administration Sequence Diagram.

After the beneficiary identity verification is completed the vaccine QR Code is scanned to extract relevant vaccine information stored on the blockchain, such as the smart contract address and the vaccine lot ID and monitoring transportation conditions. After the vaccine was administrated, a two-step locking mechanism is employed to mark the vaccine on the blockchain using the signatures of the doctor and the beneficiary. In the blockchain, the vaccine is marked as administrated using the *Hash(PI)* of the beneficiary that had received the vaccine.

The smart contract managing this process is presented in Figure 7. First, the beneficiary validation using QR code information and blockchain registered data is done (lines 4-10). This is done on the chain by offering as input the two hashes (the hash of the PI, and the hash of the SK) and checking the obtained root hash against the chain stored beneficiary registry. Both the beneficiary and the doctor must acknowledge the administration on the chain (lines 11-22) by signing the associated blockchain transaction. When both signatures are registered, the vaccine lot size is decremented and the association between the beneficiary *Hash(PI)* and the vaccine lot is registered (lines 19-20).

---

**Smart Contract:** Safe Vaccine Administration

1:    **State:**
2:       MAP (**bytes** vaccine_LID, MAP (**bytes** hashPI, MAP (**string** role, **address** actor)) **_administrationSignatures**
3:       MAP (**bytes** hashPI, **bytes** vaccine_LID) **_administratedVaccines**
4:    **Function** *CheckBeneficiaryIdentity*
5:       **Input:** hashPI, hashSecret, beneficiaryAddress
6:       **Output: validationStatus**
7:       **Begin:**

```
8:          hashBeneficiary ← keccak256 (concat(hashPI, hashSecret))
9:          validationStatus = _registeredRequests(hashBeneficiary) equal to beneficiaryAddress
10:     End
11: Function SignAdministratedVaccine
12:     Input: msg.sender, vaccineLotId, hashPI
13:     Begin:
14:         IF msg.sender is doctor THEN _administrationSignatures[vaccineLotId][hashPI][DOCTOR] = msg.sender
15:         ELSE IF msg.sender is beneficiary THEN _administrationSignatures[vaccineLotId][hashPI][BENEFI] = msg.sender
16:         IF _administrationSignatures[vaccineLotId][hashPI][DOCTOR] exists
17:                         && _administrationSignatures[vaccineLotId][hashPI][BENEF] exists
18:         THEN
19:             _ administratedVaccines[hashPI] = vaccineLotId
20:             _ vaccineLots[vaccineLotId].size = _ vaccineLots[vaccineLotId].size -1
21:         END IF
22:     End
```

**Figure 7.** Smart contract for safe vaccine administration.

Any beneficiary that has received a vaccine can register feedback and the eventual side effects encountered (see Figure 8). By registering the side effects directly on the chain, the possibility of having the information censored is made impossible. The beneficiary will sign a blockchain transaction that is authenticated and authorized (line 6) and verified against the vaccine administrated from the specified lot (line 7) and correlated with reports of the other beneficiaries. If so, the side effect registered is stored on-chain (line 8). Once the side effect is registered by the beneficiary, the information is stored as an immutable log, thus any attempt of third parties to alter it will be unsuccessful.

```
Smart Contract: Vaccine Potential Side Effects Reporting
1:  State:
2:      MAP (bytes vaccine_LID , MAP( bytes hashPI, string description)) _sideEffects
3:  Function RegisterSideEffect
4:      Input: hashPI, hashSecret, vaccineLotId, sideEffect
5:      Begin:
6:          Requires CheckBeneficiaryIdentity (hashPI, hashSecret, msg.sender) is valid
7:          Requires _ administratedVaccines[hashPI] is vaccineLotId
8:          _sideEffects[vaccineLotId][hashPI] = sideEffect
9:      End
```

**Figure 8.** Smart contract for side effect reporting.

## 4. Evaluation results

To evaluate our blockchain based system we considered a setup for the COVID-19 vaccine using the rules for safe the transportation and storage presented in [10]. We have considered that the distribution company delivers the vaccine lots to the interested medical care unit, using transportation freezers that must reach their destination in maximum 10 days, and the vaccine should be kept at a temperature between -80 and -60 Celsius degrees. Once reaching the destination, the medical unit will transfer the vaccine lots into the storage freezers where the vaccines can be deposited up to 5 days at a temperature between 2-8 degrees Celsius.

*4.1. COVID-19 vaccine distribution tracking*

A prototype has been tested on a public Ethereum test network, Ropsten [43]. The results obtained for each on chain operation can be validated on the Etherscan [43] considering the transaction hash value reported in the next sub-sections. They are describing operations executed on-chain and the associated blockchain transactions receipt is presented, highlighting the actor signing the transaction, the executed call and the transaction costs in gas.

### 4.1.1. Actors and rules registration in the immunization program

We start by deploying the smart contract used to register the main actors of the immunization campaign on the blockchain (see Table 2). The vaccine producer must use its own Ethereum address to sign the transaction that triggers the smart contract deployment.

**Table 2.** Blockchain transactions for smart contract deployment.

| Operation | Contract Deployment |
| --- | --- |
| Signing Address | 0xD2796dE988975DD292e8aC981c4011B23E801DCd *(Vaccine Issuer)* |
| Transaction Receipt | transaction hash: 0xaed1e3267afa17f22d5a125cda449945407c2485b299a8a15d9c0ff2beec6738<br>from: 0xD2796dE988975DD292e8aC981c4011B23E801DCd<br>to: VaccineRegistry.(constructor)<br>transaction: cost 2327309 gas<br>input: 0x608...40033 |

Next, the vaccine producer must sign transactions for registering the recipient medical care units as receiver of the future vaccine lots. For this a medical unit administrator must be registered on chain. In a similar manner the doctors that will administer the vaccines are registered (see Table 3).

**Table 3.** Blockchain transactions for registering immunization camping actors in this case a doctor.

| Operation | Register Doctors |
| --- | --- |
| Signing Address | 0xD2796dE988975DD292e8aC981c4011B23E801DCd *(Vaccine Issuer)* |
| Transaction Receipt | transaction hash: 0x563ddcd6fba93dbb3c05d6cb6b366a9289efcf8dd209d5516dbee66e86a8bbc6<br>from: 0xD2796dE988975DD292e8aC981c4011B23E801DCd<br>to: VaccineRegistry.registerDoctor(address) 0x536798D9D1f0507C1a8600d9A475d410a90D5A0A<br>transaction cost: 43798 gas<br>input: 0x699...26efa<br>**decoded input**  { "address doctor": "0xF3A1846C82c74EA5D5d32a9BB8759A8093C26eFa" } |

The vaccine producer must register the rules for safe transportation and storage of the vaccines. In the depicted scenario two transactions must be registered: one configuring the rules for transportation and one configuring the rules for storage (see Table 4).

**Table 4.** Blockchain transactions for registering the vaccine safe distribution rules.

| Operation | Register Tracking Rules |
| --- | --- |
| Signing Address | 0xD2796dE988975DD292e8aC981c4011B23E801DCd *(Vaccine Issuer)* |
| Rule | The vaccine should be transported at a temperature between -80 and -60 Celsius degrees for maximum 10 days. |
| Transaction Receipt **(Transportation Rule)** | transaction hash: 0xbc8b46345096d4a69faccd7300cf3450f1fc66d1e238236d511dce1771aacf9f<br>from: 0xD2796dE988975DD292e8aC981c4011B23E801DCd<br>to: VaccineRegistry.registerTrackingRules(string, int32, int32, uint256)<br>                                       0x536798D9D1f0507C1a8600d9A475d410a90D5A0A<br>transaction cost: 216219 gas<br>input: 0x4a2...00000<br>**decoded input**  {<br>        "string rule": "transport-temperature",<br>        "int32 minValue": -80,<br>        "int32 maxValue": -60,<br>        "uint256 timeDelta": {"type": "BigNumber", "hex": "0x337f9800" } } |
| Rule | Vaccine should be stored at a temperature between 2 and 8 Celsius degrees for maximum 5 days. |
| Transaction Receipt **(Storage Rule)** | transaction hash: 0xc5aa8f61103701da629ab3b5628e79ed8bf9729720d6df0e1ae1888f0e9711ec <br> from: 0xD2796dE988975DD292e8aC981c4011B23E801DCd <br> to: VaccineRegistry.registerTrackingRules(string, int32, int32, uint256)                                         0x536798D9D1f0507C1a8600d9A475d410a90D5A0A <br> transaction cost: 200595 gas <br> input: 0x4a2...26500 <br> **decoded input** { <br>      "string rule": "medicalunit-storage-temperature", <br>      "int32 minValue": 2, <br>      "int32 maxValue": 8, <br>      "uint256 timeDelta": { "type": "BigNumber", "hex": "0x19bfcc00" } } |

In a similar manner, the smart devices (freezers) that are responsible to store or transport the vaccine lots should be registered by associating one or more rules defined by the vaccine issuer to the freezer, so that the real time data feeds can be checked against these rules automatically using the smart contracts (see Table 5).

**Table 5.** Blockchain transactions for registering the freezing devices.

| Operation | Register Freezers |
|---|---|
| Signing Address | 0xD2796dE988975DD292e8aC981c4011B23E801DCd *(Vaccine Issuer)* |
| Transaction Receipt *(Transport Freezer)* | transaction hash: 0x1da71f023340fba695beeedf91ad447c7a41a726cf3296c3eb71e59fd360621e <br> from: 0xD2796dE988975DD292e8aC981c4011B23E801DCd <br> to: VaccineRegistry.registerFreezerAndRules(address,string)                                      0x536798D9D1f0507C1a8600d9A475d410a90D5A0A <br> transaction cost: 46581 gas <br> input: 0xe2d...00000 <br> **decoded input** { <br>      "address freezer": "0xA53503C7901D09358F161eC8Ec8d442d0976B9cD", <br>      "string rule": "transport-temperature" } |
| Transaction Receipt *(Storage Freezer)* | transaction hash: 0x8d84c84d19995b9904e3a8bb9ebadea0fdd989f0388c92a3c303e1e8fde459bb <br> from: 0xD2796dE988975DD292e8aC981c4011B23E801DCd <br> to: VaccineRegistry.registerFreezerAndRules(address,string)                                      0x536798D9D1f0507C1a8600d9A475d410a90D5A0A <br> transaction cost: 46701 gas <br> input: 0xe2d...26500 <br> **decoded input** { <br>      "address freezer": "0xDDb54C6fbB74a5b638EF014f7435426C46424642", <br>      "string rule": "medicalunit-storage-temperature" } |

At any point after the deployment of the smart contract, any beneficiary can subscribe on the waiting list for the vaccine. This is possible by issuing and signing a transaction on chain. The address of the signing beneficiary (*msg.sender*) will be stored in the waiting list on chain. Upon subscription the vaccine beneficiary must also provide a hash of his/her personal information (see Table 6).

**Table 6.** Blockchain transactions for registering the vaccine beneficiary

| Operation | Subscribe Beneficiary |
|---|---|
| Signing Address | 0xFfb4b11D94CFbbBA8665f7682D4d3B76261EAacC *(Beneficiary)* |
| Raw Personal Data | PI: 20-10563145-8 <br> Secret: my-super-secret |
| Hashed-Personal Data | Hash (PI): 0xa3f6550e5420ddda304a6b22772eb70b48ada3c7eb14648e321bb65387c8cfab <br> Hash (Secret): 0x820371900007448f4a8d909327870ece84168bf90f1de8dddc0b6c7473c44b40 |
| Patient Hash | Merkle Root: 0xfe08609620228b43d9eb80125dfab7a1686e9c3cd7ea5326aa1c5abf7e689b87 |
| Transaction Receipt | transaction hash: 0x76e5a604b3ca803e7738947dbc8616435d5fa410208e382a6d51b58d86b0374c<br>from: 0xFfb4b11D94CFbbBA8665f7682D4d3B76261EAacC<br>to: VaccineRegistry.registerPatient(bytes32) 0x536798D9D1f0507C1a8600d9A475d410a90D5A0A<br>transaction cost: 84808 gas<br>input: 0x8eb...89b87<br>**decoded input** {<br>    "bytes32 patientHash":<br>    "0xfe08609620228b43d9eb80125dfab7a1686e9c3cd7ea5326aa1c5abf7e689b87" } |

### 4.1.2. Vaccine tracking and administration

Once the vaccine is ready the producer can register the vaccine lots on the blockchain system specifying the number of vaccine samples in a lot and the vaccine lot ID (see Table 7).

**Table 7.** Blockchain transactions for registering the vaccine batches.

| Operation | Register Vaccine Lot |
|---|---|
| Signing Address | 0xD2796dE988975DD292e8aC981c4011B23E801DCd *(Vaccine Issuer)* |
| Transaction Receipt | transaction hash: 0x04ec1bc2f6a584810213b1521a515dc17e6f69ecda2e453ac1b554465cfa01b2<br>from: 0xD2796dE988975DD292e8aC981c4011B23E801DCd<br>to: VaccineRegistry.registerVaccineLot(bytes32, uint256)<br>                                            0x536798D9D1f0507C1a8600d9A475d410a90D5A0A<br>transaction cost: 64255 gas<br>Input: 0x0a1...000c8<br>**decoded input** {<br>    "bytes32 vaccineLotId":<br>        "0xd7adb300b4c0d0f79bbb9195e3f9513b49caf8d14383062b2032d5656b13c5b5",<br>    "uint256 samples": { "type": "BigNumber", "hex": "0xc8" } } |

Next, the vaccine lot is associated with one of the registered freezers. Each time a vaccine transfer is carried out on the distribution chain it will be marked on the blockchain chain by updating the freezer associated with the vaccine lot (see Table 8).

**Table 8.** Blockchain transactions for registering the vaccine lot freezing device for distribution.

| Operation | Register Vaccine Lot for Transportation |
|---|---|
| Signing Address | 0xD2796dE988975DD292e8aC981c4011B23E801DCd *(Vaccine Issuer)* |
| Transaction Receipt | transaction hash: 0xf840c0653a1190a875825377afb7feacf130e7fc32d81bde887ae71ef388a7c9<br>from: 0xD2796dE988975DD292e8aC981c4011B23E801DCd<br>to: VaccineRegistry.updateVaccineFreezer(bytes32, address, address)<br>                                            0x536798D9D1f0507C1a8600d9A475d410a90D5A0A<br>transaction cost: 68106 gas<br>input: 0xf63...6b9cd<br>**decoded input** {<br>    "bytes32 vaccineLotId":<br>        "0xd7adb300b4c0d0f79bbb9195e3f9513b49caf8d14383062b2032d5656b13c5b5",<br>    "address freezerDeviceNew": "0xA53503C7901D09358F161eC8Ec8d442d0976B9cD",<br>    "address freezerDeviceOld": "0xA53503C7901D09358F161eC8Ec8d442d0976B9cD" } |

During transportation, the freezer will register on the chain the values received from the associated sensors (in our case the values are received from temperature sensors). They will be registered in an immutable manner as transactions on the blockchain system allowing the future audit of the vaccine distribution conditions against the producer rules (see Table 9).

**Table 9.** Blockchain transactions for distribution tracking.

| Operation | Vaccine distribution tracking |
|---|---|
| Signing Address | 0xA53503C7901D09358F161eC8Ec8d442d0976B9cD *(Transport Freezer)* |
| Transaction Receipt | transaction hash: 0x0338487f1a35aa763bde543d39159e26875ab70742303cc583fe3ce638279998<br>from: 0xA53503C7901D09358F161eC8Ec8d442d0976B9cD<br>to: VaccineRegistry.monitor (bytes32, string, int32) 0x536798D9D1f0507C1a8600d9A475d410a90D5A0A<br>transaction cost: 160278 gas<br>input: 0x524...00000<br>**decoded input** {<br>    "bytes32 vaccineLotId": "0xd7adb300b4c0d0f79bbb9195e3f9513b49caf8d14383062b2032d5656b13c5b5",<br>    "string rule": "transport-temperature",<br>    "int32 monitoredValue": -70} |
| Transaction Receipt | transaction hash: 0xc816e94acd01eebfd524f49b16e910f53910c91eda4445102c15a7653c0f8668<br>from: 0xA53503C7901D09358F161eC8Ec8d442d0976B9cD<br>to: VaccineRegistry.monitor (bytes32, string, int32) 0x536798D9D1f0507C1a8600d9A475d410a90D5A0A<br>transaction cost: 129212 gas<br>input: 0x524...00000<br>**decoded input** {<br>    "bytes32 vaccineLotId": "0xd7adb300b4c0d0f79bbb9195e3f9513b49caf8d14383062b2032d5656b13c5b5",<br>    "string rule": "transport-temperature",<br>    "int32 monitoredValue": -55 }<br>**decoded output** -<br>logs [ { "from": "0x536798D9D1f0507C1a8600d9A475d410a90D5A0A", "topic": "0x285c0714a79fa669178437acfc343777469b86a755b376b8143ad3087951d959",<br>**"event"**: *"BrokenRule"*, **"args"**: { "0": "transport-temperature", "1": "0xd7adb300b4c0d0f79bbb9195e3f9513b49caf8d14383062b2032d5656b13c5b5", "2": "-55", "3": "1607426813",<br>**"rule": "transport-temperature",**<br>**"vaccineLot": "0xd7adb300b4c0d0f79bbb9195e3f9513b49caf8d14383062b2032d5656b13c5b5",**<br>**"value": "-55",**<br>**"time": "1607426813" } } ]** |
| …. | … more monitored values |

After reaching the medical care center, the beneficiaries will be scheduled for having the vaccine administered. Any vaccine beneficiary once reaching the doctor's office will have to provide the personal identification information QR code. The doctor will scan the QR code which will offer information about the beneficiary and the transaction hash proving that the subscribing vaccination list transaction has been mined (Table 10).

**Table 10.** Beneficiary QR code and data registered on blockchain.

| QR Code | Information stored |
|---|---|
| 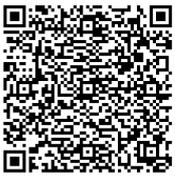 | PI: 20-10563145-8<br>Hash_Secret: 0x820371900007448f4a8d909327870ece84168bf90f1de8dddc0b6c7473c44b40<br>Contract: 0x536798D9D1f0507C1a8600d9A475d410a90D5A0A<br>TX_Hash: 0x76e5a604b3ca803e7738947dbc8616435d5fa410208e382a6d51b58d86b0374c |

Using this information, the patient will be verified against the waiting list registered on chain (see Table 11).

**Table 11.** Checking beneficiary subscription to the vaccine waiting list.

| Operation | Check Patient Subscription |
|---|---|
| Signing Address | 0xF3A1846C82c74EA5D5d32a9BB8759A8093C26eFa *(Doctor)* |
| Call (Calls are not mined they are only queries to check the state) | from 0xF3A1846C82c74EA5D5d32a9BB8759A8093C26eFa<br>to VaccineRegistry.checkPatientRegistration(bytes32, bytes32, address)<br>0x536798D9D1f0507C1a8600d9A475d410a90D5A0A<br>Input: 0x615...eaacc<br>**decoded input** {<br>   "bytes32 hashPI":<br>      "0xa3f6550e5420ddda304a6b22772eb70b48ada3c7eb14648e321bb65387c8cfab",<br>   "bytes32 hashSecret":<br>      "0x820371900007448f4a8d909327870ece84168bf90f1de8dddc0b6c7473c44b40",<br>   "address patient": "0xFfb4b11D94CFbbBA8665f7682D4d3B76261EAacC"}<br>**decoded output** {"0": "bool: true"} |

Once validated, the doctor will check the vaccine sample using its QR code associated, specifying the lot Id and the address of the blockchain smart contract where the tracking information is registered (see Table 12).

**Table 12.** Vaccine identification QR code.

| QR Code | Information stored |
|---|---|
| 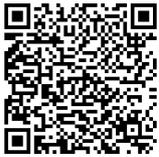 | V_ID<br>0xd7adb300b4c0d0f79bbb9195e3f9513b49caf8d14383062b2032d5656b13c5b5<br>Contract<br>0x536798D9D1f0507C1a8600d9A475d410a90D5A0A |

Using this information, the patient can verify the tracking information and whether the vaccine lot was correctly transported (see Table 13).

**Table 13.** Check vaccine distribution tracking information.

| Operation | Check Vaccine History |
|---|---|
| Signing Address | 0xFfb4b11D94CFbbBA8665f7682D4d3B76261EAacC *(Beneficiary)* |
| Call (Calls are not mined they are only queries to check the state) | From: 0xFfb4b11D94CFbbBA8665f7682D4d3B76261EAacC<br>To: VaccineRegistry.checkVaccineLotHistory(bytes32)<br>    0x536798D9D1f0507C1a8600d9A475d410a90D5A0A<br>Input: 0xfb7...3c5b5<br>**decoded input** {<br>   "bytes32 vaccineLotId":<br>      "0xd7adb300b4c0d0f79bbb9195e3f9513b49caf8d14383062b2032d5656b13c5b5" }<br>**decoded output** {<br>   "0": "tuple (address, string,int32,uint256,bool)[]:<br>      0xA53503C7901D09358F161eC8Ec8d442d0976B9cD,<br>        transport-temperature, -70, 1607426735, **true**,<br>      0xA53503C7901D09358F161eC8Ec8d442d0976B9cD,<br>        transport-temperature, -55, 1607426813, **false**,<br>      0xDDb54C6fbB74a5b638EF014f7435426C46424642,<br>        medicalunit-storage-temperature, 5, 1607427059, **true**,<br>      0xDDb54C6fbB74a5b638EF014f7435426C46424642,<br>        medicalunit-storage-temperature, 10, 1607427155, **false**"} |

Once the vaccine is administered, a multi-signature is required on the chain for updating the vaccine lot size and marking one vaccine sample as administered. The signatures are expected from both the receiving beneficiary and the doctor which has administrated the vaccine (see Table 14).

**Table 14.** Signing the acknowledgement of vaccine administration.

| Operation | Signing |
| --- | --- |
| Signing Address | 0xFfb4b11D94CFbbBA8665f7682D4d3B76261EAacC *(Beneficiary)* |
| Transaction Receipt | transaction hash: 0x24813e343ccb5f4b5916367c5337385c494ff3e576a5e1ffbcd5c9ee1f424508<br>from: 0xFfb4b11D94CFbbBA8665f7682D4d3B76261EAacC<br>to: VaccineRegistry.signAdministeredVaccine(bytes32, bytes32) 0x536798D9D1f0507C1a8600d9A475d410a90D5A0A<br>transaction cost: 49401 gas<br>input: 0x4df...8cfab<br>**decoded input** {<br>"bytes32 vaccineLotId": "0xd7adb300b4c0d0f79bbb9195e3f9513b49caf8d14383062b2032d5656b13c5b5", "bytes32 hashPI": "0xa3f6550e5420ddda304a6b22772eb70b48ada3c7eb14648e321bb65387c8cfab" } |
| Signing Address | 0xF3A1846C82c74EA5D5d32a9BB8759A8093C26eFa *(Doctor)* |
| Transaction Receipt | transaction hash: 0xf8d21694b2008cd123fa89899ed5b3a330df4460c53f25d3d15e8b1cdef76d4e<br>from: 0xF3A1846C82c74EA5D5d32a9BB8759A8093C26eFa<br>to: VaccineRegistry.signAdministeredVaccine(bytes32,bytes32) 0x536798D9D1f0507C1a8600d9A475d410a90D5A0A<br>transaction cost: 74528 gas<br>input: 0x4df...8cfab<br>**decoded input** {<br>"bytes32 vaccineLotId": "0xd7adb300b4c0d0f79bbb9195e3f9513b49caf8d14383062b2032d5656b13c5b5", "bytes32 hashPI": "0xa3f6550e5420ddda304a6b22772eb70b48ada3c7eb14648e321bb65387c8cfab" } |

Once both signatures are received, the vaccine is considered successfully administered and the vaccine lot size decreases by one. This can be verified on chain by any participant. After receiving the vaccine, the patient can register voluntarily any side effect he is feeling (see Table 15).

**Table 15.** Potential side effects reporting.

| Operation | Register Side effect |
| --- | --- |
| Signing Address | 0xFfb4b11D94CFbbBA8665f7682D4d3B76261EAacC *(Beneficiary)* |
| Transaction Receipt | transaction hash:0x7ed30d2cd822d854bc36ba666c657e5029e7f6d72c1c67202aabff459a658a13<br>from: 0xFfb4b11D94CFbbBA8665f7682D4d3B76261EAacC<br>to: VaccineRegistry.registerSideEffect(bytes32, bytes32, bytes32, string) 0x536798D9D1f0507C1a8600d9A475d410a90D5A0A<br>transaction cost: 48073 gas<br>input: 0x56f...00000<br>**decoded input** {<br>"bytes32 hashPI": "0xa3f6550e5420ddda304a6b22772eb70b48ada3c7eb14648e321bb65387c8cfab",<br>"bytes32 hashSecret": "0x820371900007448f4a8d909327870ece84168bf90f1de8dddc0b6c7473c44b40",<br>"bytes32 vaccineLotId": "0xd7adb300b4c0d0f79bbb9195e3f9513b49caf8d14383062b2032d5656b13c5b5", "string sideEffect": "Dizziness, Nausea" } |

## 4.2. Throughput and Scalability

The integration of monitored data feed of vaccine distribution condition directly on blockchain may feature high costs associated with the cumulated blockchain transactions and poor scalability determined by the block mining periodicity and gas limit imposed per block. To deal with these issues we have integrated a pre-processing step at the edge level (physical device level) that is responsible to receive all the data from one sensor deployed on the vaccine distribution network and determine for an interval the most significant values are registered on the chain. In this case, the most significant values are the lowest and the highest temperature values registered in that interval. This solution had been presented in detail in one of our previously published research [44].

For scalability experiments, we have considered the setup and restrictions in terms of gas limits (approximately 12 000 000 per block) and mining periodicity (i.e. 15 seconds) of the public Ethereum main network. In our blockchain-based system case, we obtain a transaction cost of approximately 140 000 gas for registering the monitored temperature of the freezing devices concerning vaccine safe delivery rules and transaction throughput of approximately 85 transactions per block.

The transaction mining results are presented in Figure 9 considering an interval of one hour in which each device will have to store two temperature monitoring transactions (one for the minimum registered value and one for the maximum registered value). Up to 10 000 freezers that are exclusively signing transactions on the blockchain network can be efficiently managed without creating a bottleneck. Being a linear relationship one can easily change the interval for reporting monitored temperature transactions to accommodate more transportation devices.

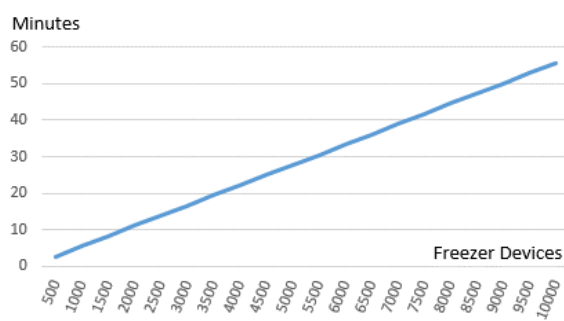

**Figure 9.** Mining time for vaccine distribution temperature tracking transactions.

**5. Conclusions**

In this paper we presented a blockchain-based system for transparent tracing of COVID-19 vaccine registration, storage and delivery and side effects self-reporting. Blockchain is used to offer data immutability, transparency and correctness of beneficiary registration for vaccination, eliminating identity thefts and impersonations. The tracking and monitoring of vaccine distribution against the producer defined rules for safe manipulation is done using decentralized smart contracts. Also, a blockchain solution is proposed for vaccine administration and transparent and tamper proof self-reporting of side effects, person identification and vaccine association.

The results provided for an Ethereum based implementation show the feasibility of our proposed solution in terms of immutable actors and rules registration, decentralized vaccine distribution monitoring and finally administration and potential side effects self-reporting. The proposed system manages to successfully address all relevant aspects we had identified for the success of a monitoring campaign: (i) increase the efficiency and transparency of COVID-19 vaccine distribution assuring the traceability and the rigorous audit of the storage and delivery conditions (ii) assure the transparency and correctness in the registration and management of the waiting list for immunization and (iii) provide a transparent and public reporting system of potential side effects.